\documentclass[a4paper]{jpconf}
\usepackage{graphicx}
\usepackage{color}
\usepackage{latexsym}
\usepackage{amsfonts}

\newcommand{\be}{\begin{eqnarray}}
\newcommand{\ee}{\end{eqnarray}}

\begin{document}

\title{Testing the Kerr black hole hypothesis with {\sc relxill\_nk}}

\author{Cosimo Bambi}

\address{Center for Field Theory and Particle Physics and Department of Physics, Fudan University, 200438 Shanghai, China}
\address{Theoretical Astrophysics, Eberhard-Karls Universit\"at T\"ubingen, 72076 T\"ubingen, Germany}

\ead{bambi@fudan.edu.cn}

\begin{abstract}
Astrophysical black hole candidates are thought to be the Kerr black holes of general relativity. However, macroscopic deviations from the Kerr background are predicted by a number of scenarios beyond Einstein's gravity. X-ray reflection spectroscopy can be a powerful tool to probe the strong gravity region of these objects and test the Kerr black hole hypothesis. Here I briefly review the state of the art of this line of research and I present some constraints on possible deviations from the Kerr metric obtained with the new X-ray reflection model {\sc relxill\_nk} and XMM-Newton, NuSTAR, and Swift data of the supermassive black hole in 1H0707-495.
\end{abstract}


\section{Introduction}

The theory of general relativity was proposed by Einstein in 1915. In the past sixty years, the theory has been extensively tested in weak gravitational fields, mainly with experiments in the Solar System and observations of binary pulsars, and current data well agree with the theoretical predictions~\cite{will}. However, there are a number of scenarios beyond Einstein's gravity that have the same predictions for weak fields and differ when gravity becomes strong. The ideal laboratory to test strong gravity is the spacetime around astrophysical black holes~\cite{book}.

In 4-dimensional Einstein's gravity, the only stationary, asymptotically flat, regular on and outside the event horizon (i.e. without singularities and closed time-like curves), vacuum solution is the Kerr metric~\cite{hair}, which is completely characterized by two parameters, namely the black hole mass $M$ and the black hole spin angular momentum $J$. In Boyer-Lindquist coordinates, the line element reads ($G_{\rm N} = c = 1$)
\be\label{eq-k}
ds^2 &=& - \left(1 - \frac{2 M r}{\Sigma} \right) \, dt^2
+ \frac{\Sigma}{\Delta} \, dr^2 + \Sigma \, d\theta^2 \nonumber\\
&& + \left( r^2 + a^2 + \frac{2 a^2 M r \sin^2\theta}{\Sigma}\right) \, d\phi^2 
- \frac{4 a M r \sin^2\theta}{\Sigma} \, dt \, d\phi \, ,
\ee
where $a = J/M$, $\Sigma = r^2 + a^2 \cos^2\theta$, and $\Delta = r^2 - 2 M r + a^2$. $M$ and $J$ cannot be arbitrary and must satisfy the condition for the existence of the even horizon $|a_*| \le 1$, where $a_* = J/M^2$ is the dimensionless spin parameter.

The spacetime metric around astrophysical black holes formed from the gravitational collapse of massive bodies is expected to be well approximated by the Kerr solution~\cite{book}. Initial deviations from the Kerr metric are quickly radiated away with emission of gravitational waves after the formation of the horizon. The equilibrium electric charge is extremely small and completely negligible for the spacetime geometry. The presence of accretion disks can be usually ignored. In the end, macroscopic deviations from the Kerr metric are only possible in the presence of new physics.


\section{How can we test the Kerr black hole hypothesis?}

There are two main approaches to test the Kerr nature of astrophysical black holes: with electromagnetic radiation~\cite{review,review2} and with gravitational waves~\cite{review-gw}. Each method has its own advantages and disadvantages, and the two techniques are actually complementary, because deviations from standard predictions may lead to observational effects in one of the two approaches and not in the other and vice versa.

The electromagnetic approach is sensitive to the motion of the gas particles in the accretion disk and to the propagation of the photons from the point of emission in the strong gravity region to the point of detection in the flat faraway region. There are two natural strategies to test the Kerr black hole hypothesis with electromagnetic radiation~\cite{book}. In the so-called top-down approach, we consider some alternative theory of gravity in which black holes are not described by the Kerr metric and we check whether astrophysical data prefer the Kerr metric of Einstein's gravity or the non-Kerr metric of that specific alternative theory of gravity. This strategy has two main problems. First, there are a large number of alternative theories of gravity, and none seems to be more motivated than others, so we should repeat the analysis for every theory. Second, rotating black hole solutions in alternative theories of gravity are known only in quite exceptional cases, while the non-rotating or slow-rotating solutions are not very useful to test astrophysical black holes because the spin plays an important role in the properties of the electromagnetic radiation.

In the bottom-up approach, we employ a phenomenological test-metric in which possible deviations from the Kerr solution are quantified by one or more ``deformation parameters''. We measure the values of these deformation parameters with astrophysical data and we check whether they vanish, as it is required to recover the Kerr metric of Einstein's gravity.


\section{X-ray reflection spectroscopy}

X-ray reflection spectroscopy refers to the study of the reflection spectrum of accretion disks around black holes (see, e.g., \cite{book} and references therein). Within the disk-corona model, a black hole is surrounded by a geometrically thin and optically thick accretion disk. The disk radiates like a blackbody locally and as a multi-color blackbody when integrated radially. The ``corona'' is a hotter, usually optically thin, electron cloud enshrouding the disk. Its exact geometry is currently unknown: it may be the base of the jet, the atmosphere above the disk, some kind of accretion flow between the disk and the central black holes, etc. Because of inverse Compton scattering of thermal photons from the disk off free electrons in the corona, the latter becomes an X-ray source with a power-law spectrum. Some X-ray photons from the corona hit the accretion disk, producing a reflection component with some fluorescent emission line.

The most prominent feature of the reflection spectrum is usually the iron K$\alpha$ line, which is at 6.4~keV in the case of neutral and weakly ionized iron and shifts up to 6.97~keV for H-like iron ions. While this line is very narrow in the rest-frame of the emitter, the one observed in the spectrum of black holes is broad and skewed due to special and general relativistic effects (gravitational redshift, Doppler boosting, light bending) occurring in the strong gravity region. In the presence of the correct astrophysical model, the analysis of the iron K$\alpha$ line and of the whole reflection spectrum can be a powerful tool to probe the spacetime around the black hole. The technique was originally developed assuming the Kerr metric to measure the black hole spin parameter $a_*$~\cite{i1,i2,i3}. More recently, it was extended to generic metrics to test the Kerr black hole hypothesis~\cite{j1,j2,j3,j4,j4b,j5,j6}.


\section{\sc relxill\_nk}

{\sc relxill} is currently the most advanced model to describe the X-ray reflection spectrum of accretion disks in the Kerr spacetime~\cite{rel1,rel2,rel3}. Its extension to non-Kerr spacetimes is {\sc relxill\_nk} and was presented in~\cite{relnk}. {\sc relxill\_nk} currently employ the Johannsen metric with the deformation parameter $\alpha_{13}$~\cite{j-m}, but can easily be extended to any stationary, axisymmetric, and asymptotically flat black hole spacetime. In Boyer-Lindquist coordinates, the line element reads
\be
ds^2 &=& - \frac{\Sigma \left(\Sigma - 2 M r \right)}{A^2} \, dt^2
+ \frac{\Sigma}{\Delta} \, dr^2 + \Sigma \, d\theta^2
+ \frac{\left[ \left(r^2 + a^2\right)^2 \left(1 + \delta\right)^2 
- a^2 \Delta \sin^2\theta\right] 
\Sigma \sin^2\theta}{A^2} \, d\phi^2
\nonumber\\ &&
- \frac{2 a \left[ 2 M r + \delta \left(r^2 + a^2\right) \right] 
\Sigma \sin^2\theta}{A^2} \, dt \, d\phi \, ,
\ee
where $A = \Sigma + \delta \left(r^2 + a^2\right)$ and $\delta = \alpha_{13} \left(M/r\right)^3$. For $\alpha_{13} = 0$, we exactly recover the Kerr metric in~(\ref{eq-k}). $\alpha_{13}$ cannot be arbitrary; it must satisfy the condition
\be\label{eq-bound}
\alpha_{13} \ge - \left( 1 + \sqrt{1 - a^2_*} \right)^3 \, ,
\ee
in order to have a regular exterior region (no singularities or closed time-like curves).


\section{Observational constraints from 1H0707-495}

The study reported in Ref.~\cite{zheng} is the first attempt to test the Kerr nature of astrophysical black holes with real X-ray data. As source, we considered the supermassive black hole in the galaxy 1H0707-495: its spectrum is characterized by significant edge features, which are commonly interpreted as an extremely strong reflection component. This suggests that such a source can be a good candidate for testing the Kerr black hole hypothesis with {\sc relxill\_nk}.

\begin{figure}[t]
\begin{center}
\includegraphics[type=pdf,ext=.pdf,read=.pdf,width=7.9cm]{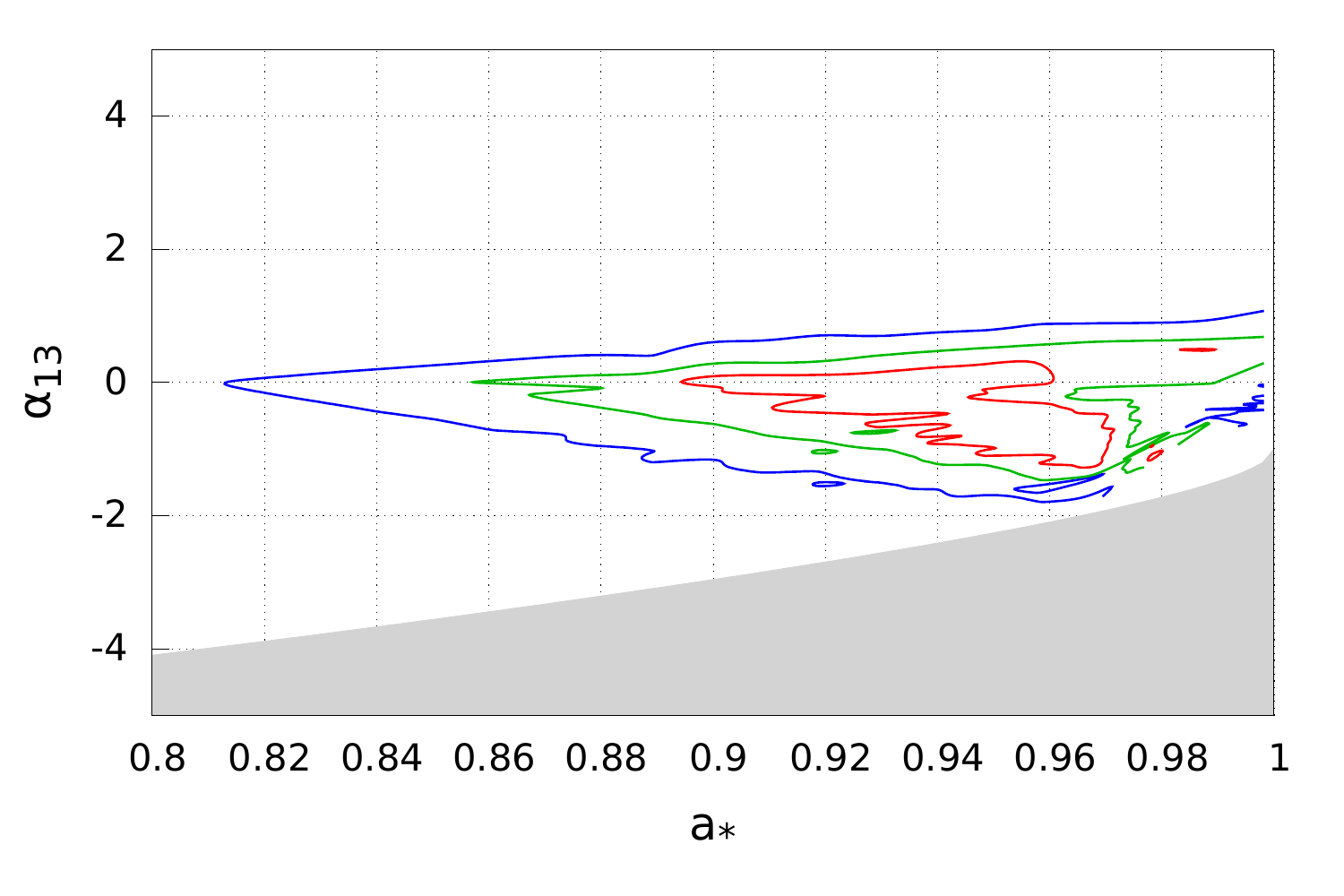}
\includegraphics[type=pdf,ext=.pdf,read=.pdf,width=7.9cm]{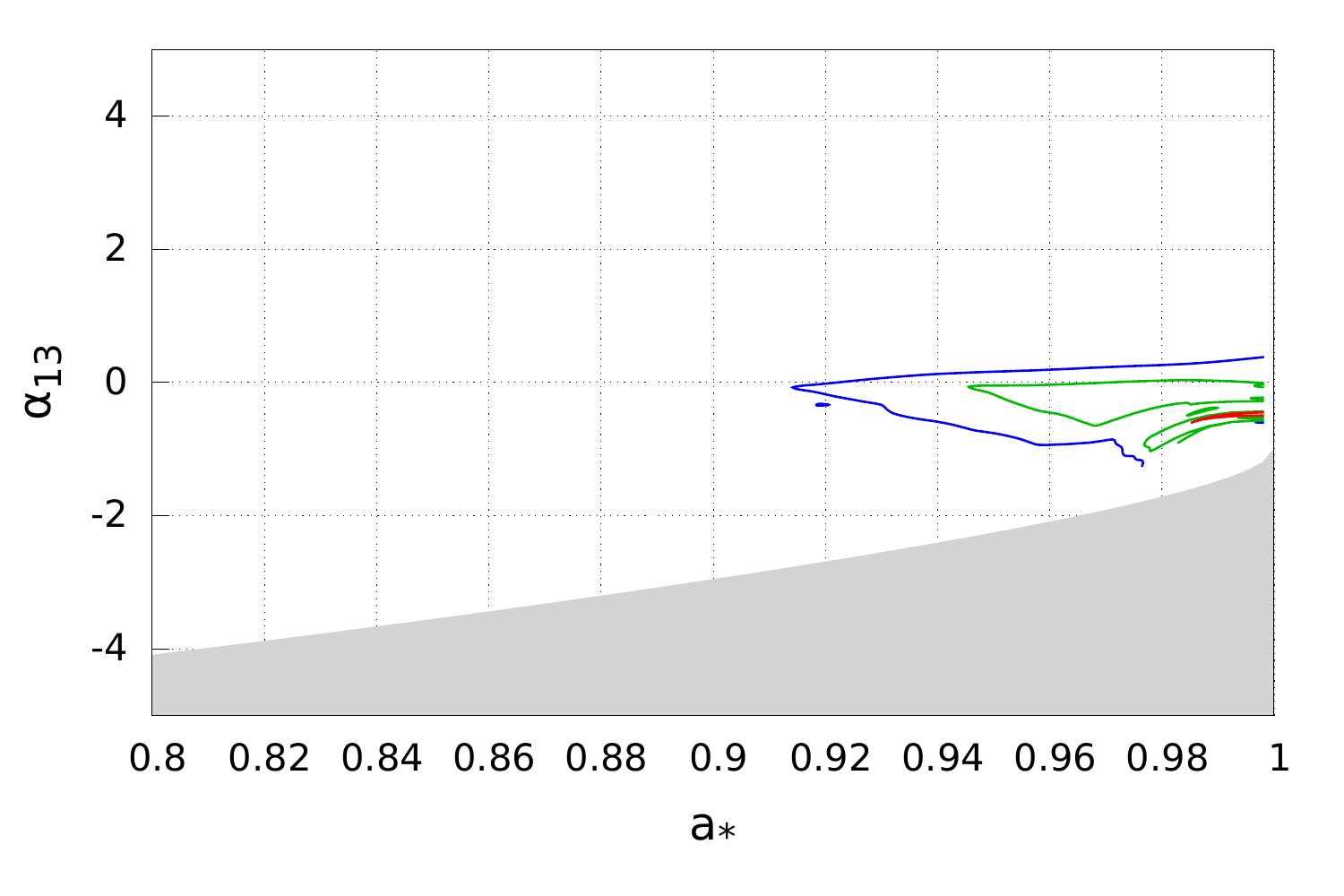}
\end{center}
\vspace{-0.5cm}
\caption{Constraints on the spin parameter $a_*$ and the Johannsen deformation parameter $\alpha_{13}$ from the XMM-Newton data of 2011: Model~1 (left panel) and Model~2 (right panel). The red, green, and blue lines indicate, respectively, the 68\%, 90\%, and 99\% confidence level curves for two relevant parameters. The grayed region is outside the range prescribed for $\alpha_{13}$ in Eq.~(\ref{eq-bound}) and therefore is ignored in our study. From Ref.~\cite{zheng}. \label{f-xmm}}
\vspace{0.5cm}
\begin{center}
\includegraphics[type=pdf,ext=.pdf,read=.pdf,width=7.9cm]{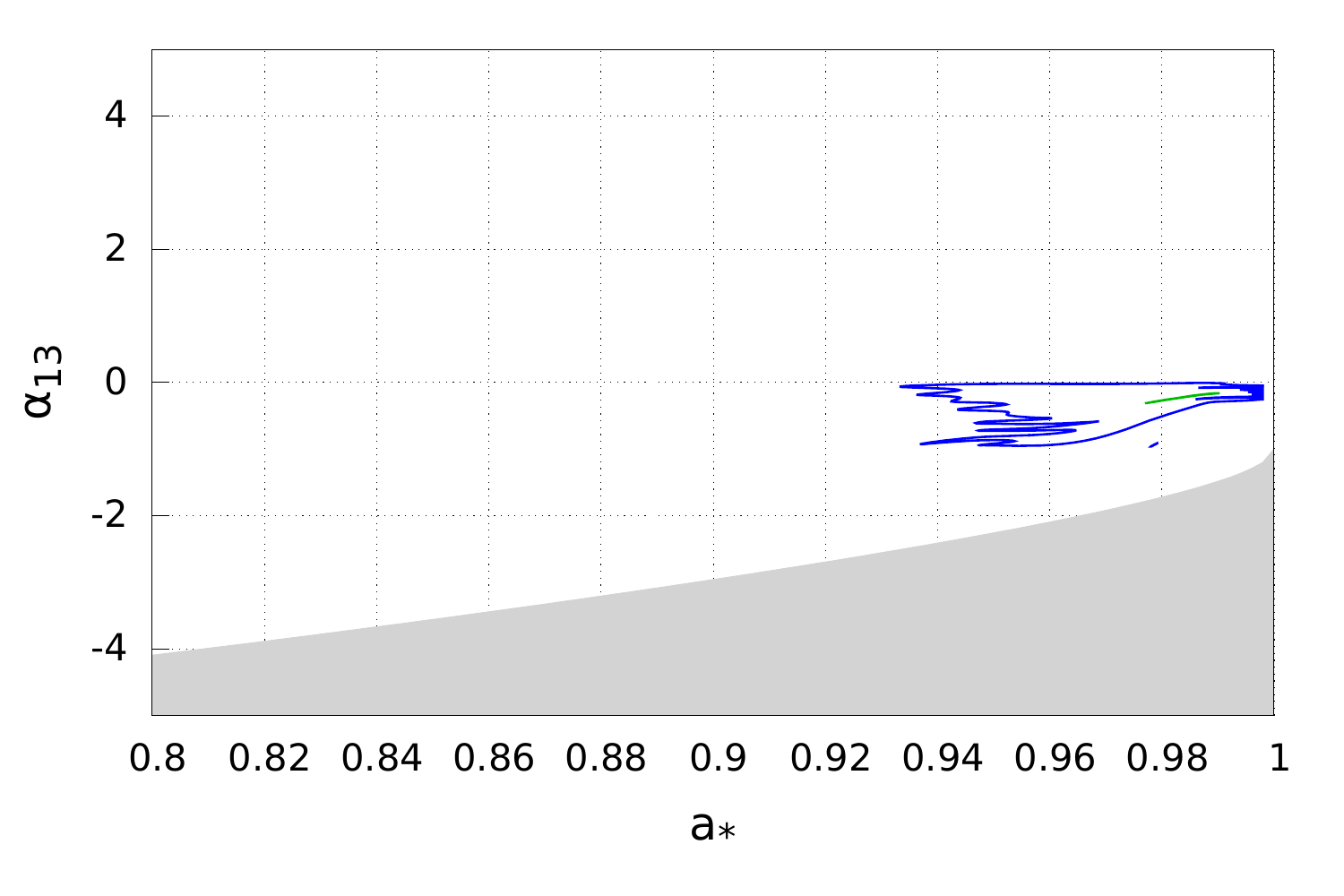}
\end{center}
\vspace{-0.5cm}
\caption{As in Fig.~\ref{f-xmm} for Model~3 and the data NuSTAR+Swift. The 68\% confidence level curve is too thin to be plotted. From Ref.~\cite{zheng}. \label{f-nustar}}
\end{figure}

There are several observations of 1H0707-495 with XMM-Newton, NuSTAR, and Swift. For XMM-Newton, we considered the 98~ks observation of 2011, which corresponds to the lowest flux state of 1H0707-495 and has been investigated by several authors. We fitted the data with two models
\be\label{eq-m1m2}
&& {\rm Model~1 : \,\,\, {\sc TBabs*(relxill\_nk+diskbb)}} \, , \nonumber\\
&& {\rm Model~2 : \,\,\, {\sc TBabs*(relxill\_nk+relxill\_nk)}} \, . 
\ee
{\sc TBabs} takes the galactic dust absorption into account. {\sc relxill\_nk} is the disk's reflection spectrum. {\sc diskbb} is the thermal component of a Newtonian disk. See~\cite{zheng} and reference therein for more details.
Fig.~\ref{f-xmm} shows the constraints on the spin parameter $a_*$ and the deformation parameter $\alpha_{13}$ from Model~1 (left panel) and Model~2 (right panel). The red, green, and blue lines indicate, respectively, the 68\%, 90\%, and 99\% confidence level curves.

Concerning NuSTAR and Swift data, there are three separated observations of 1H0707-495 with NuSTAR in 2014 and each observation has a simultaneous snapshot of Swift. In our study we excluded the second Swift observation because it was taken during an anomaly period of this mission. We employed the following model (for more details, see Ref.~\cite{zheng})
\be\label{eq-m3}
&& {\rm Model~3 : \,\,\, {\sc TBabs*relxill\_nk}} \, . 
\ee
Fig.~\ref{f-nustar} shows the constraints on $a_*$ and $\alpha_{13}$ obtained from the NuSTAR+Swift data. The green and blue lines indicate, respectively, the 90\% and 99\% confidence level curves, while we do not report the 68\% confidence level curve because it is too thin to be plotted.

\section{Opportunities with future X-ray missions}

\begin{figure*}[t]
\begin{center}
\includegraphics[type=pdf,ext=.pdf,read=.pdf,width=7.9cm]{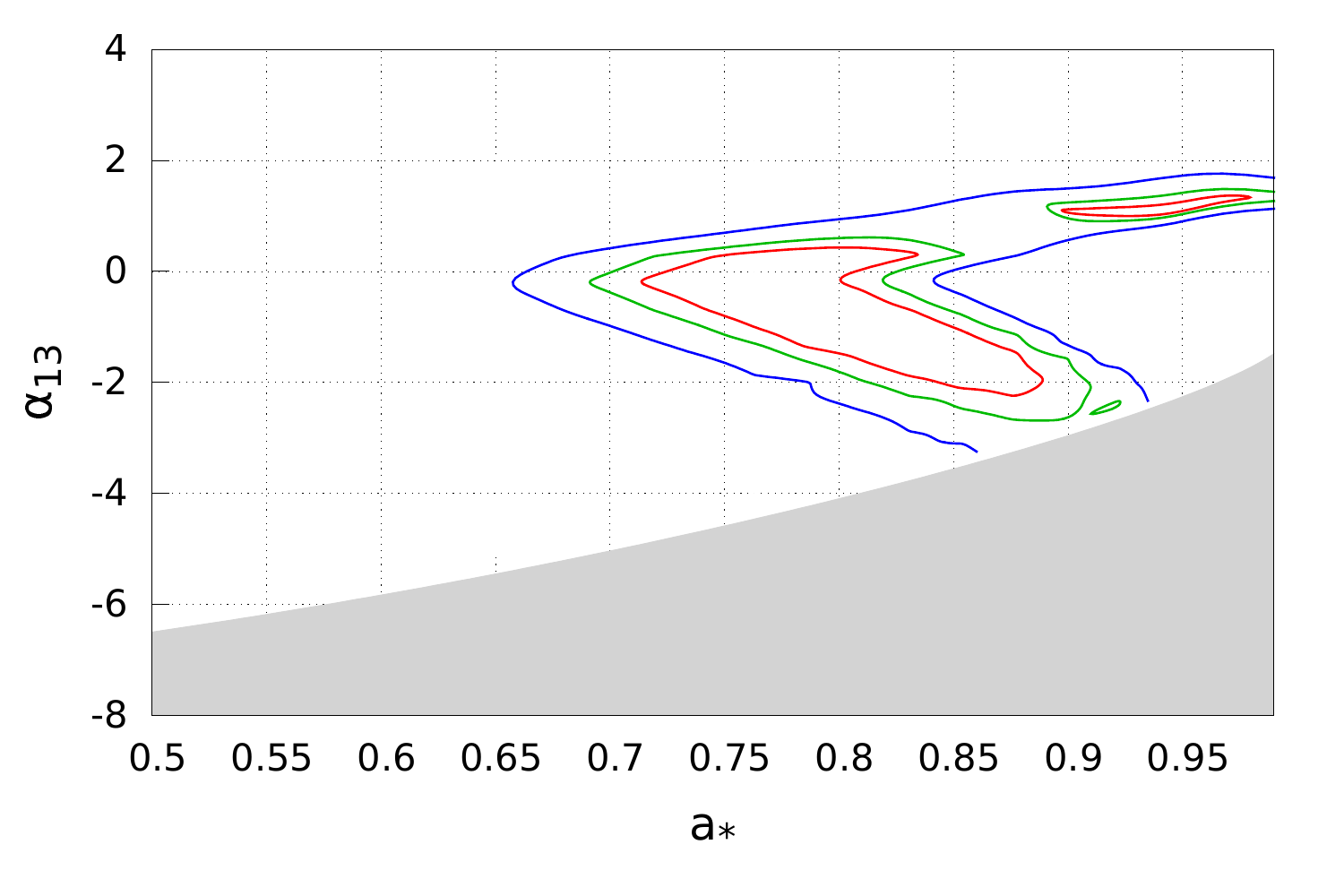}
\includegraphics[type=pdf,ext=.pdf,read=.pdf,width=7.9cm]{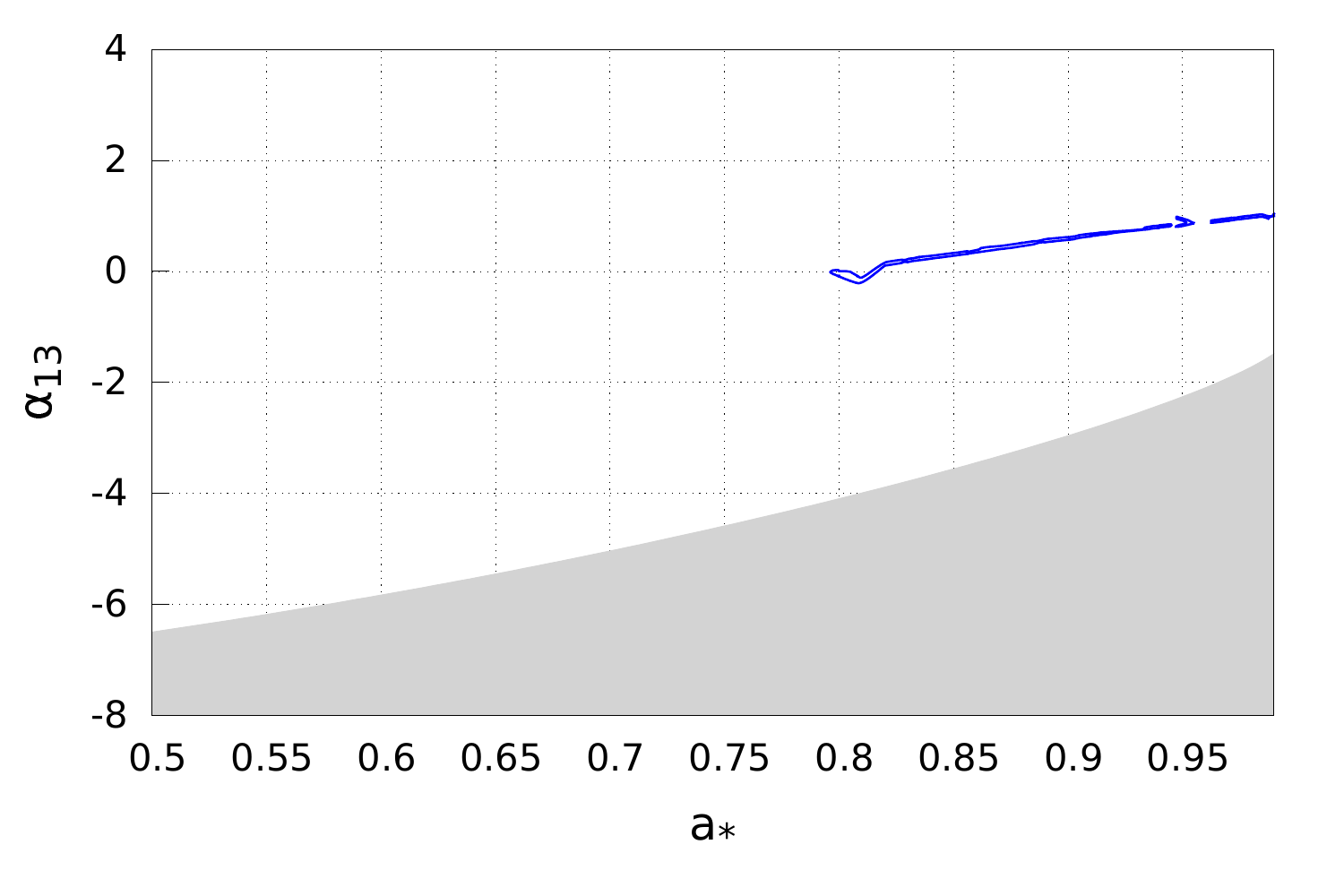}
\end{center}
\vspace{-0.5cm}
\caption{68\%, 90\%, and 99\% confidence level curves for the spin parameter $a_*$ and the deformation parameter $\alpha_{13}$ from a simulated observation of a bright stellar-mass black hole with NuSTAR (left panel) and LAD/eXTP (right panel). The spacetime metric of the simulation has $\alpha_{13} = 0$ (Kerr) and $a_* = 0.8$; the viewing angle is $i = 30^\circ$. From Ref.~\cite{relnk}. \label{f-sim1}}
\vspace{0.5cm}
\begin{center}
\includegraphics[type=pdf,ext=.pdf,read=.pdf,width=7.9cm]{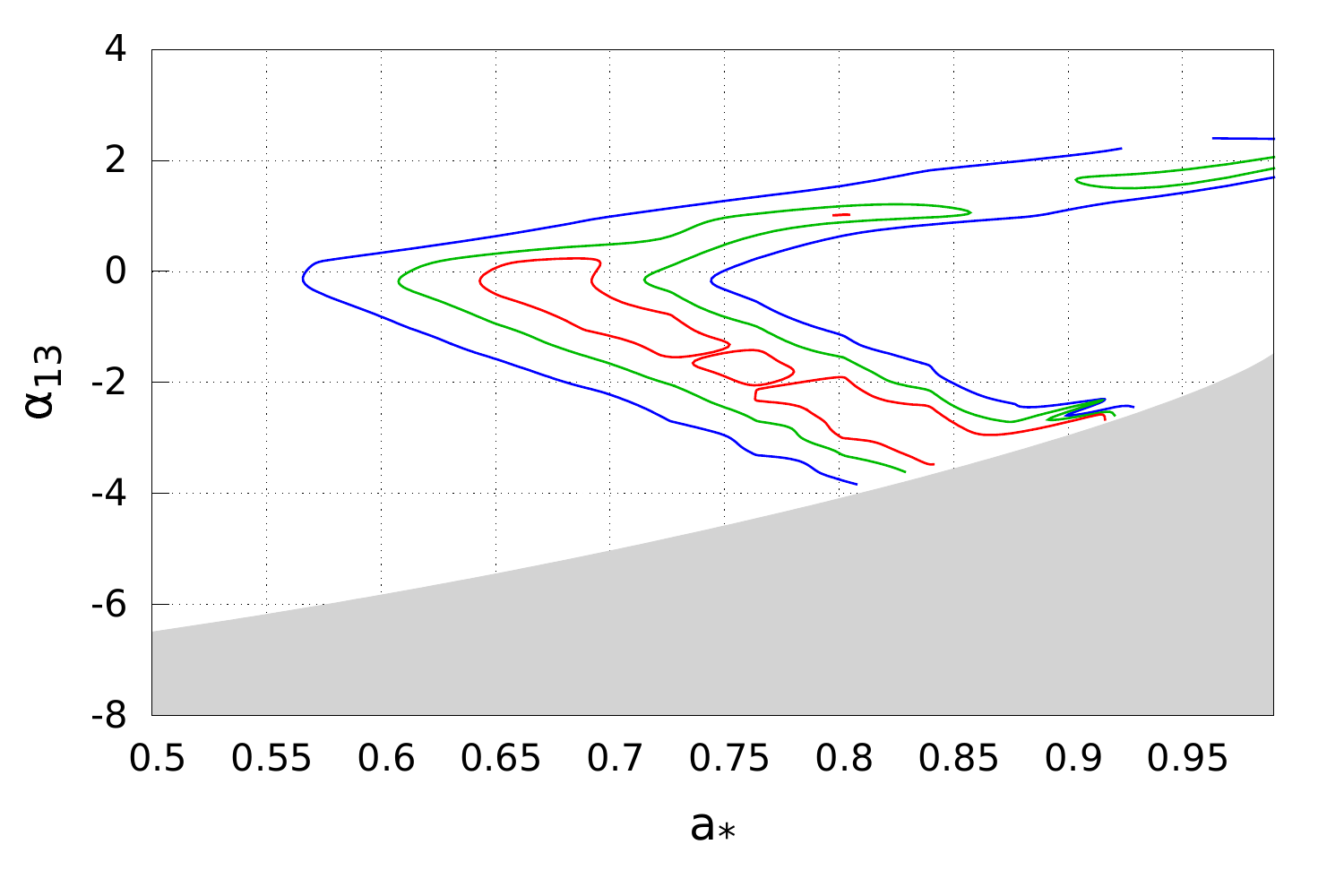}
\includegraphics[type=pdf,ext=.pdf,read=.pdf,width=7.9cm]{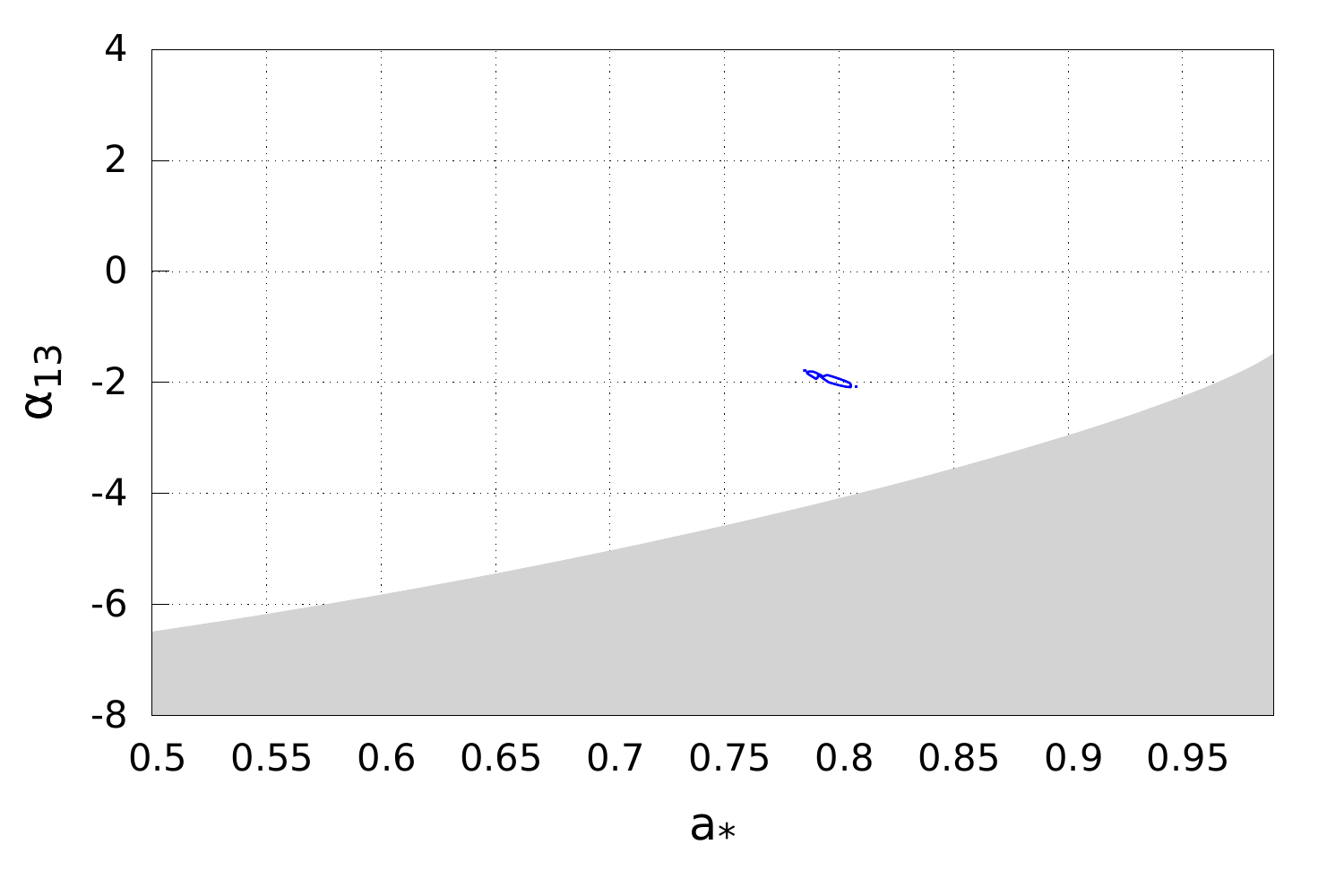}
\end{center}
\vspace{-0.5cm}
\caption{As in Fig.~\ref{f-sim1} for $\alpha_{13} = -2$, $a_* = 0.8$, and $i = 30^\circ$. From Ref.~\cite{relnk}. \label{f-sim2}}
\vspace{0.5cm}
\begin{center}
\includegraphics[type=pdf,ext=.pdf,read=.pdf,width=7.9cm]{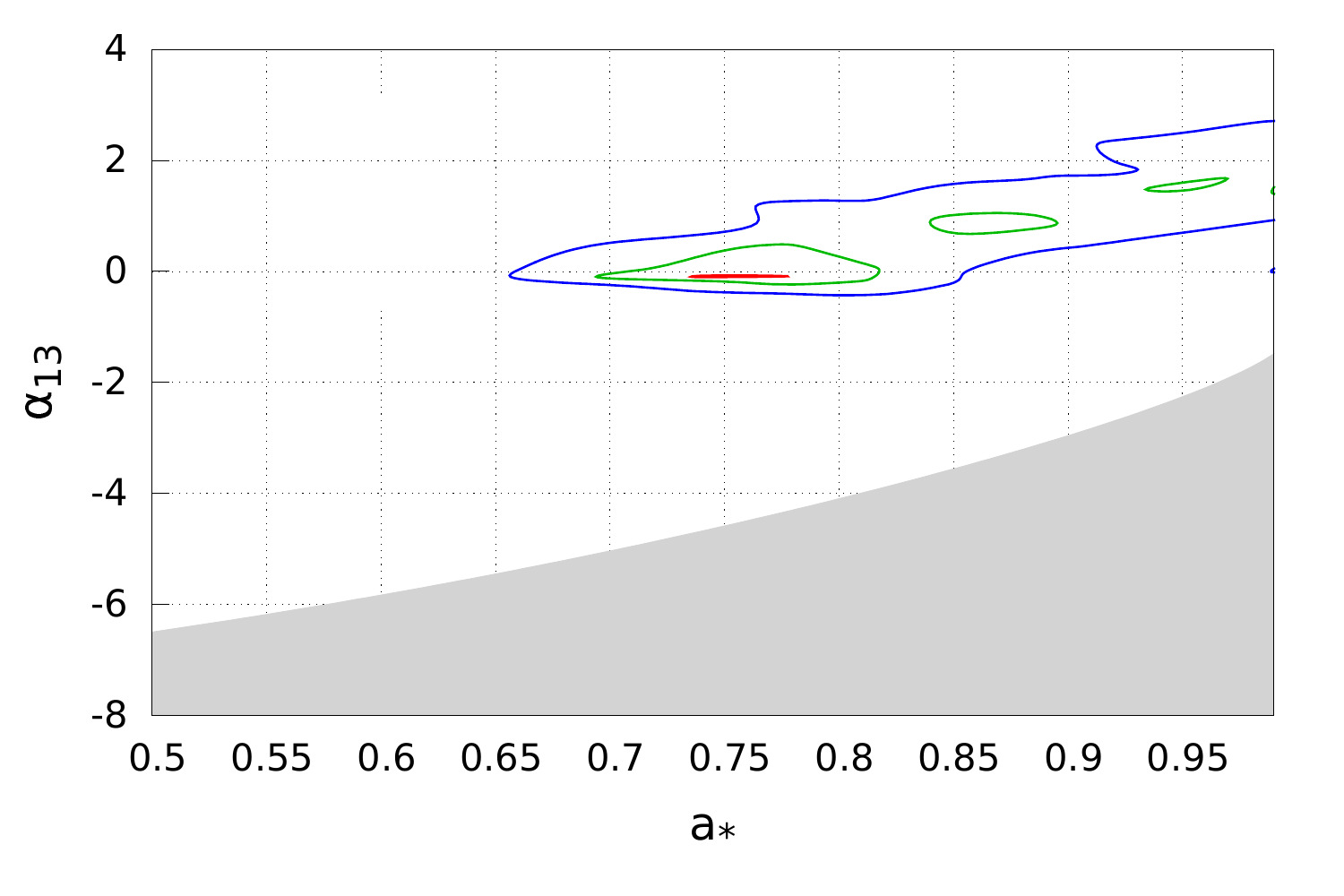}
\includegraphics[type=pdf,ext=.pdf,read=.pdf,width=7.9cm]{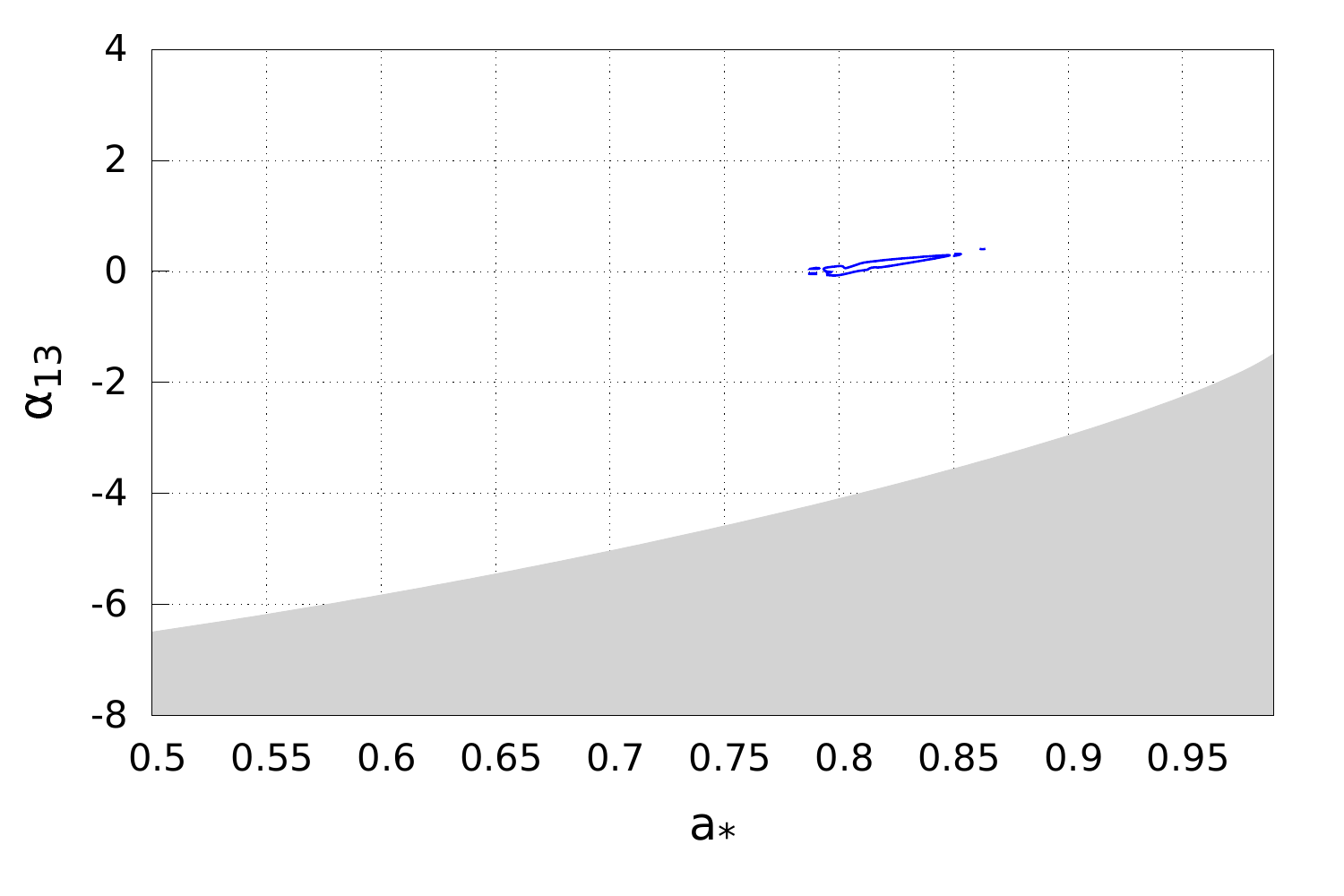}
\end{center}
\vspace{-0.5cm}
\caption{As in Fig.~\ref{f-sim1} for $\alpha_{13} = 0$ (Kerr), $a_* = 0.8$, and $i = 80^\circ$. From Ref.~\cite{relnk}. \label{f-sim3}}
\end{figure*}

In Ref.~\cite{relnk}, simulations were performed to test the capabilities of {\sc relxill\_nk} in analyzing observations from NuSTAR (as an example of a present instrument) and LAD/eXTP~\cite{snzhang} (as an example of the next generation of X-ray missions). The resulting constraints are shown in Figs.~\ref{f-sim1}, \ref{f-sim2}, and \ref{f-sim3}. In every figure, the left panel is for NuSTAR, the right panel for LAD/eXTP. In Fig.~\ref{f-sim1}, the reference model is a Kerr black hole with the spin parameter $a_* = 0.8$ and the inclination angle $i = 30^\circ$. In Fig.~\ref{f-sim2}, we considered a non-Kerr black hole with the deformation parameter $\alpha_{13} = -2$, the spin parameter $a_* = 0.8$, and the inclination angle $i = 30^\circ$. Fig.~\ref{f-sim3} shows the constraints obtained from a simulated observation of a Kerr black hole with the spin parameter $a_* = 0.8$ and the inclination angle $i = 80^\circ$.

The constrains obtained simulating data with NuSTAR in Ref.~\cite{relnk} are comparable with those obtained in the analysis of 1H0707-495 in Ref.~\cite{zheng}. We can thus expect that the constraining power of the next generation of X-ray mission is significantly better, as we find in our simulations.

\section{Concluding remarks}

{\sc relxill\_nk} is the extension of the X-ray reflection model {\sc relxill} to probe the metric around astrophysical black holes and test the Kerr black hole hypothesis using X-ray reflection spectroscopy~\cite{relnk}. In Ref.~\cite{zheng}, we employed for the first time the new model to analyze XMM-Newton, NuSTAR, and Swift data of the supermassive black hole in 1H0707-495. Our results are shown in Figs.~\ref{f-xmm} and \ref{f-nustar}, and are consistent with the assumption that the metric around the supermassive black hole in 1H0707-495 is described by the Kerr solution, as expected in Einstein's gravity. We are currently working to test the Kerr metric with other sources (both stellar-mass and supermassive black hole candidates) as well as different kinds of deformations from the Kerr solution. Simulations show that significantly better constraints can be obtained with the next generation of X-ray missions~\cite{relnk}.


\ack

This work was supported by the National Natural Science Foundation of China (Grant No.~U1531117), Fudan University (Grant No.~IDH1512060), and the Alexander von Humboldt Foundation.


\end{document}